**Temperature dependence of $^7$Li NMR relaxation rates in Li$_3$InCl$_6$, Li$_3$YCl$_6$, Li$_{1.48}$Al$_{0.48}$Ge$_{1.52}$(PO$_4$)$_3$ and LiPS$_5$Cl**

*Darshan Chalise $^{1,2*}$, Carlos Juarez-Yescas$^{2,3}$, Beniamin Zahiri$^{2,4}$, Paul V. Braun$^{2,3,4,5,6}$, David G. Cahill$^{1,2,4*}$*

$^1$*Department of Physics, University of Illinois at Urbana-Champaign, Urbana, IL, 61801, USA*

$^2$*Materials Research Laboratory, University of Illinois at Urbana-Champaign, Urbana, IL, 61801, USA*

$^3$*Department of Chemistry, University of Illinois Urbana-Champaign, Urbana, IL 61801, USA*

$^4$ *Department of Materials Science and Engineering, University of Illinois at Urbana-Champaign, Urbana, IL, 61801, USA*

$^5$*Department of Chemical and Biomolecular Engineering, University of Illinois Urbana-Champaign, Urbana, IL 61801, USA*

$^6$*Beckman Institute for Advanced Science and Technology, University of Illinois Urbana-Champaign, Urbana, IL 61801, USA*

**Abstract**

Inorganic solid-state battery electrolytes show high ionic conductivities and enable the fabrication of all solid-state batteries. In this work, we present the temperature dependence of spin-lattice relaxation time ($T_1$), spin-spin relaxation time ($T_2$), and resonance linewidth ($\varDelta v$) of the $^7$Li nuclear magnetic resonance (NMR) for four solid-state battery electrolytes (Li$_3$InCl$_6$ (LIC), Li$_3$YCl$_6$ (LYC), Li$_{1.48}$Al$_{0.48}$Ge$_{1.52}$(PO$_4$)$_3$ (LAGP) and LiPS$_5$Cl (LPSC)) from 173 K to 403 K at a $^7$Li resonance frequency of 233 MHz, and from 253 K to 353 K at a $^7$Li resonance frequency of 291 MHz. Additionally, we measured the spin-lattice relaxation rates at an effective $^7$Li resonance frequency of 133 kHz using a spin-locking pulse sequence in the temperature range of 253 K to 353 K. In LPSC, the $^7$Li NMR relaxation is consistent with the Bloembergen-Pound-Purcell (BPP) theory of NMR relaxation of dipolar nuclei. In LIC, LYC and LAGP, the BPP theory does not describe the NMR relaxation rates for the temperature range and frequencies of our measurements. The presented NMR relaxation data assists in providing a complete picture of Li diffusion in the four solid-state battery electrolytes.

**Introduction**

Solid-state battery electrolytes overcome the traditional safety concerns of leakage and unimpeded dendrite growth in liquid state electrolytes and can be used with a lithium metal anode [1,2]. The high capacity of lithium metal anodes[3], combined with stacking efficiency due to structural integrity of the solid-state electrolytes[4] means that high volumetric and specific energy densities can be achieved in solid-state batteries. Inorganic solid-state electrolytes show significantly higher ionic conductivities compared to solid polymer electrolytes[3], and therefore make all solid-state batteries with inorganic solid electrolytes a promising next-generation technology for energy storage.

NMR spectroscopy of solid-state battery electrolytes have been used to measure the diffusion constant and its temperature dependence[2,5,6]. The diffusion constant is intrinsically linked to the ionic conductivity of the solid-state electrolyte through the Nernst-Einstein equation[2,6]. Puled field gradient (PFG) NMR allows for direct measurement of the Li diffusion constant[6–8]. However, due to small values of the diffusion constant and the short Li spin-spin relaxation time, measurements of the Li diffusion in solid-state electrolytes typically requires specialized PFG probes with gradient strength > 10 T/m[7,8]. Alternatively, the Bloembergen-Bloom-Purcell (BPP) model[9] for the dipolar interaction between two Li nuclei is used to relate the NMR relaxation rates with the correlation time of interaction between two lithium ions. The correlation time is then used to estimate Li diffusion constant through the Einstein-Smoluchowski equation[2,5,6].

In this work, we measure the spin-lattice relaxation time ($T_1$), spin-spin relaxation rate ($T_2$), resonance linewidth ($\Delta v$) and the consequent coherence time of the resonance ($T_2^* = \frac{1}{\Delta v}$) of $Li_3InCl_6$ (lithium indium chloride or LIC), $Li_3YCl_6$ (lithium yttrium chloride or LYC), $Li_{1.48}Al_{0.48}Ge_{1.52}(PO_4)_3$ (lithium aluminum germanium phosphate or LAGP) and $LiPS_5Cl$ (lithium phosphorous sulphur chloride or LPSC) as a function of temperature at 233 MHz and 291 MHz. Additionally, using spin-locking sequence[10] at 17.1T, we also measure the spin-spin relaxation time at an effective rotational frame of reference frequency of 133 kHz. The data presented here shows that while the BPP model for dipolar interaction of two nuclei is applicable for LiPS$_5$Cl at the $^7$Li nuclear resonance frequencies and temperature of our study, the BPP model does not describe the $^7$Li NMR relaxation in Li$_3$InCl$_6$, Li$_3$YCl$_6$ and Li$_{1.48}$Al$_{0.48}$Ge$_{1.52}$(PO$_4$)$_3$.

**Experimental Details**

**Material Synthesis**

Li$_3$YCl$_6$ and Li$_3$InCl$_6$ solid electrolytes (SEs) were synthesized via intermittent high energy ball milling as previously reported[11,12]. In summary, LiCl (Sigma-Aldrich) and YCl$_3$ (Alfa Aesar), or LiCl and InCl$_3$ (Alfa Aesar) were hand mixed using a mortar with a 3:1 stoichiometry. The mixture was then milled intermittently for 3 hours. In the case of Li$_3$YCl$_6$, a 10% molar excess of YCl$_3$ was added to account for loss of material due to strong adhesion to the milling reactor. Material manipulation and synthesis was carried out in an Ar glovebox (O$_2$ and H$_2$O, <0.1 ppm). Li$_6$PS$_5$Cl (NEI Corporation) and Li$_{1.48}$Al$_{0.48}$Ge$_{1.52}$(PO$_4$)$_3$ (MSE Supplies) were purchased as powders. The vendor specified average particle of Li$_6$PS$_5$Cl and Li$_{1.48}$Al$_{0.48}$Ge$_{1.52}$(PO$_4$)$_3$ were 1 $\mu$m and 500 nm respectively.

**Materials Characterization**

The X-ray diffraction (XRD) patterns of the solid-state electrolytes were collected using a Bruker D8 Advance Plus instrument at the Materials Research Lab at the University of Illinois. The XRD experiment was carried out in powder mode (Cu K$\alpha$, $\lambda$=1.5418 Å) from 10° to 80° in 2$\theta$. To protect the air and moisture sensitive electrolytes, a custom holder and a Kapton film (Chemplex) were used.

The ionic conductivity of the solid-state electrolytes (SEs) was measured via electrochemical impedance spectroscopy (EIS). The SEs were first pelletized by weighing approximately 100 mg of SE, then pressed to ~360 MPa between two Ti rods. The Ti rods also served as blocking electrodes. In the case of LAGP, it was first pelletized to ~360 MPa, then sintered at 700°C for 12 hours in air since LAGP is air stable[12]. Then, to ensure good electrical contact, the sintered LAGP pellet coated with 100 nm Au on both sides via thermal evaporation using a Kurt J. Lesker Nano36 thermal evaporator. The EIS measurements were recorded in the frequency range from 1 MHz to 100 mHz with a 10 mV ac amplitude using a Bio-Logic VMP3 impedance analyzer.

Energy dispersive X-ray fluorescence (EDXRF) measurements were performed using a Shimadzu EDXRF 7000 instrument to ensure the stoichiometry of In and Cl in $LiInCl_6$, Y and Cl in $LiYCl_6$, P and Cl and P and S in $LiPS_5Cl$, and Al and P in $Li_{1.48}Al_{0.48}Ge_{1.52}(PO_4)_3$ were as expected. The EDXRF measurements were carried out using x-rays generated by a Rh anode with acceleration voltage of 50 kV. The measured stoichiometries were determined from calibration curve method using standard samples ($InCl_3$, $YCl_3$, $PCl_5$, $PS_5$ and $Al(PO_4)_3$) using K-α x-ray emissions for Cl, P, S and Al; and L-α x-ray emissions for In and Y. The experimental uncertainty determined from repeated measurements on the standard sample was approximately 1% for each element.

NMR measurements were performed at 233 MHz in a 14.1 T Unity Inova NMR spectrometer over a temperature range of 173 K to 403 K. We used a 5 mm Varian broadband probe in the temperature range 213 K to 373 K, and a 10 mm probe for measurements below 213 K and above 373 K. We used a 17.1 T Agilent VNMRS system with a 4mm Varian solid-state probe for measurements at 291 MHz. Additionally, rotational frame of reference measurement with spin-locking sequence was performed in VNMRS system with effective resonance frequency of 133 kHz.

We calibrated the temperature of the NMR probes using the relative chemical shifts between two proton peaks of ethylene glycol for temperatures above 308 K and methanol below 308 K[13].

Spin-Lattice relaxation times ($T_1$) were measured using the inversion recovery sequence[14], spin-spin relaxation times ($T_2$) were measured using the Carr-Purcell-Meiboom-Gill (CPMG) sequence[15] and the resonance linewidths were obtained by Fourier transforming the free induction decay (FID) of the resonance. $T_{1,\delta}$ measurements at 17.1 T were performed using a spin-locking sequence[10].

All NMR data were processed in Mnova with single exponential fits for the determination of $T_1$, $T_2$ and $T_{1,\delta}$ values.

**Results**

X-ray diffraction (XRD) data verify that the correct phases were obtained in all four solid-state electrolytes, see Figure 1. XRD showed that Li$_3$InCl$_6$ crystalized in the monoclinic phase, with the $C2/m$ space group[16]. The XRD pattern for Li$_3$YCl$_6$ shows a crystalline phase indexed to the trigonal structure with the $P\bar{3}m1$ space group, in agreement with previous reports[4,12,17]. XRD also confirmed Li$_6$PS$_5$Cl has an argyrodite-type structure (space group $F\bar{4}3m$)[18] and Li$_{1.48}$Al$_{0.48}$Ge$_{1.52}$(PO$_4$)$_3$ crystallizes in the $R\bar{3}c$ space group[19].

EXDRF measurements showed a In:Cl ratio of 1:6.1 in Li$_3$InCl$_6$, a Al:P ratio of 1:6.2 in Li$_{1.48}$Al$_{0.48}$Ge$_{1.52}$(PO$_4$)$_3$, and P:Cl and P:S ratios of 1:1 and 1:4.9 in Li$_6$PS$_5$Cl. In Li$_3$YCl$_6$, a 1:5.7 ratio of Y:Cl was observed, consistent with the 10% additional YCl$_3$ added for stabilization.

The ionic conductivity was measured using electrochemical impedance spectroscopy (EIS). We derive the bulk ionic resistance ($R_b$) of the electrolytes from the Nyquist plots using a simple equivalent circuit [20,21]. We calculate the the ionic conductivity using $\sigma = \frac{t}{R_b A}$ where $t$ is the thickness and $A$ is the area of the pelletized solid-state electrolyte. The corresponding Nyquist plots and ionic conductivities are shown in Figure 2. The measured conductivities of LIC, LYC, LPSC and LAGP agree with previous reports[4,11,16,18,22]. In LAGP, the measured ionic conductivity increased after sintering, but still remained low, typical when the grain boundary resistance is high[22,23]. Larger grain boundary resistance did not affect NMR measurements as the NMR results in LAGP did not change significantly before and after sintering.

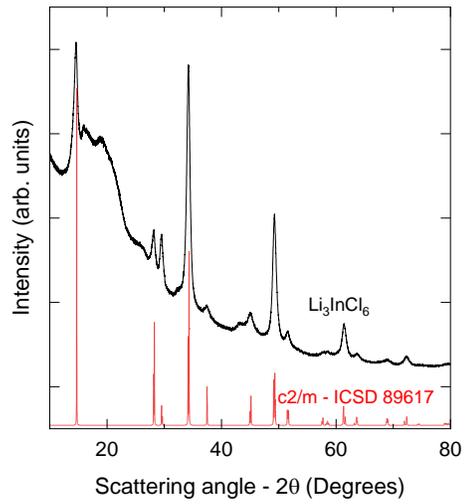
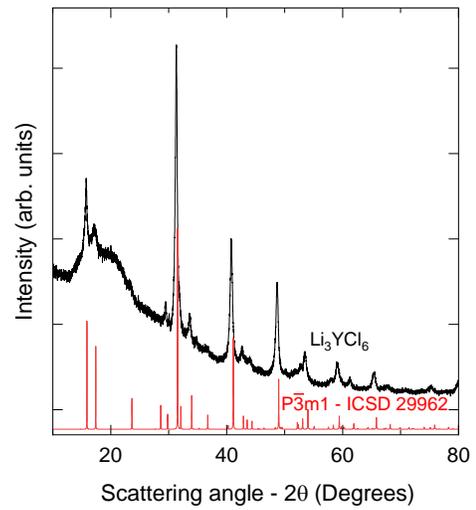
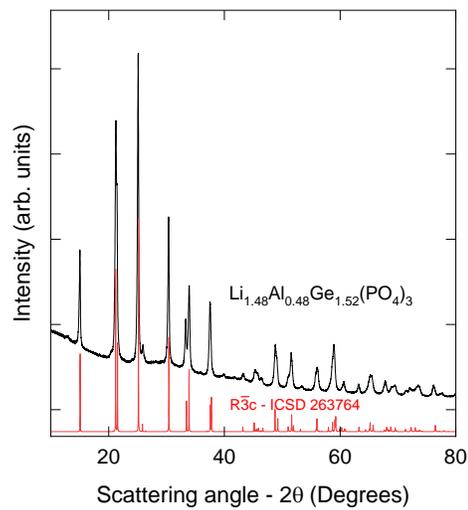
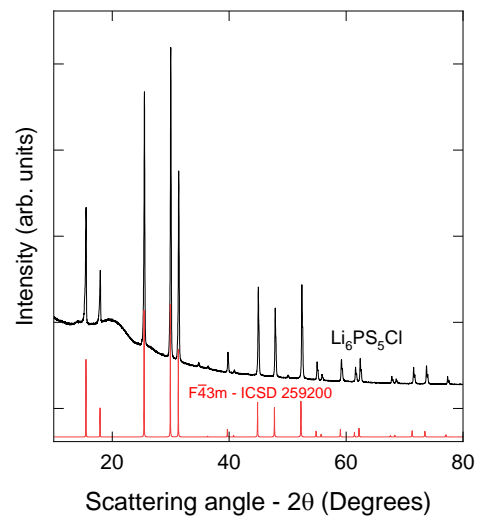

Figure 1. Powder x-ray diffraction of (a) $Li_3InCl_6$, (b) $Li_3YCl_6$, (c) $Li_{1.48}Al_{0.48}Ge_{1.52}(PO_4)_3$ and (d) $Li_6PS_5Cl$. The powder x-ray diffraction intensities collected in θ-2θ geometry using Cu Kα x-rays is displayed in black and the simulated spectra for corresponding space groups is displayed in red. The diffuse scattering background, prominent for lower scattering angles, is from Kapton layer used to prevent exposure of the electrolytes to air.

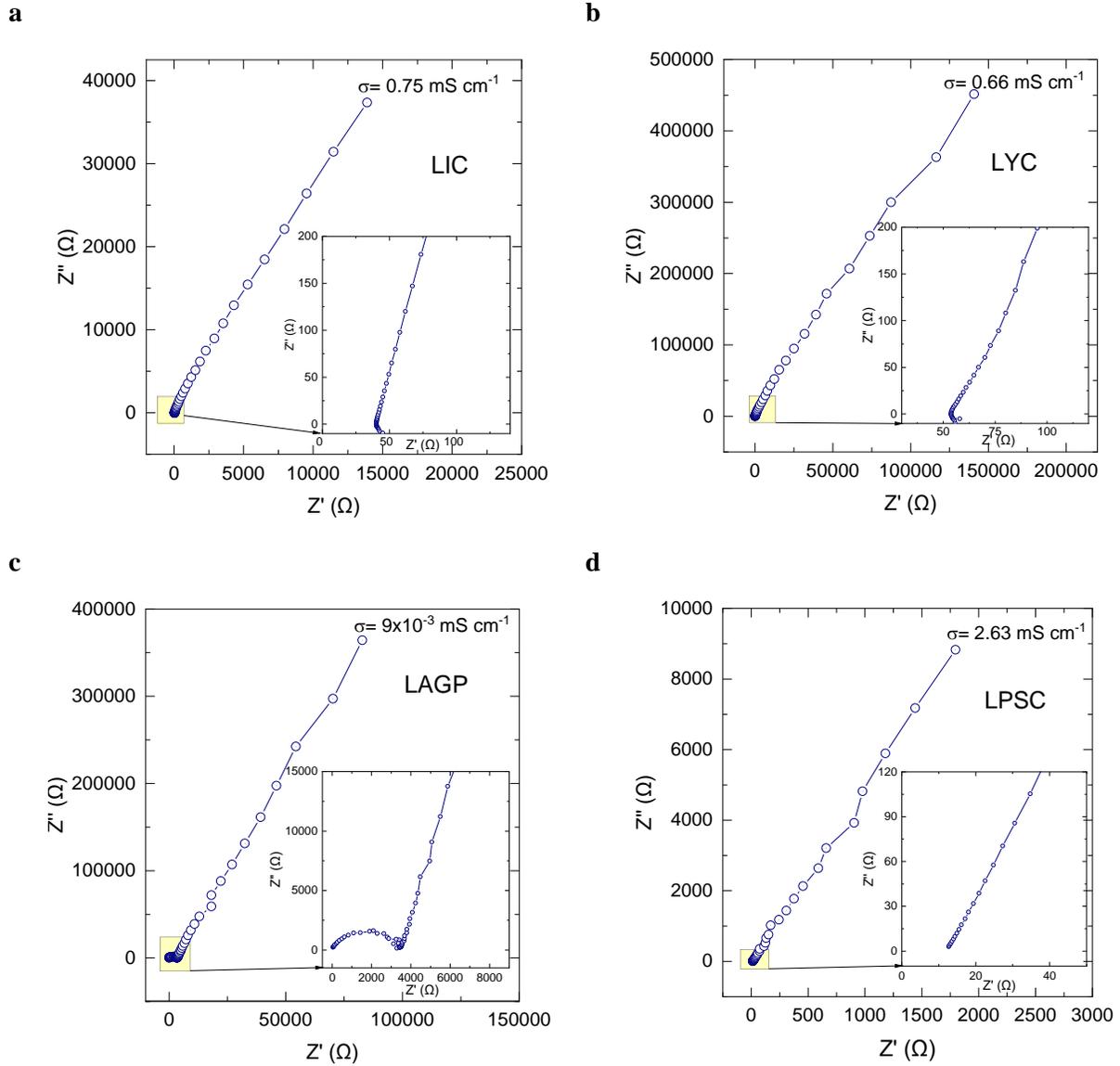

Figure 2. Electrochemical impedance spectroscopy (EIS) Nyquist plots for (a) $LiInCl_6$, (b) $LiYCl_6$, (c) $Li_{1.48}Al_{0.48}Ge_{1.52}(PO_4)_3$ and (d) $Li_6PS_5Cl$ taken between 1 MHz to 100 mHz. The behavior at higher frequencies are represented by the insets in the plots. The determined conductivities are 0.75 mS cm$^{-1}$, 0.66 mS cm$^{-1}$, $9 \times 10^{-3}$ mS cm$^{-1}$ and 2.63 mS cm$^{-1}$ for $LiInCl_6$, $LiYCl_6$, $Li_{1.48}Al_{0.48}Ge_{1.52}(PO_4)_3$ and $Li_6PS_5Cl$ respectively.

The $^7$Li NMR relaxation times $T_1$, $T_2$ and the NMR linewidth at 291 MHz and 233 MHz as well as the relaxation rate $T_{1,\delta}$ in the effective rotational frame of reference frequency of 133 kHz for $LiInCl_6$, $LiYCl_6$, $Li_{1.48}Al_{0.48}Ge_{1.52}(PO_4)_3$ and $Li_6PS_5Cl$ for different temperatures are displayed in Tables 1-4 respectively. The relaxation rates $1/T_1$, $1/T_2$, $1/T_2^*$ and $1/T_{1,\delta}$ are plotted as a function of temperatures for in Figure 3.

Table 1. NMR spin-lattice relaxation time ($T_1$), spin-spin relaxation time ($T_2$), NMR linewidth ($\Delta v$) at 291 MHz and 233 MHz and the rotational frame of reference spin-lattice relaxation time $T_{1,\delta}$ at an effective frequency of 133 kHz for $Li_3InCl_6$.

291 MHz

| T(K) | $\Delta v$ (Hz) | $T_1$ (s) | $T_2$ (ms) | $T_{1,\delta}$ (ms), $f_{eff}$ = 133 kHz |
|---|---|---|---|---|
| 353 | 967 | 0.50 | 2.99 | 5.49 |
| 343 | 1022 | 0.50 | 2.29 | 4.37 |
| 323 | 1120 | 0.55 | 2.04 | 2.77 |
| 313 | 1178 | 0.55 | 1.33 | 2.40 |
| 303 | 1124 | 0.54 | 1.20 | 2.10 |
| 293 | 1441 | 0.57 | 1.13 | 1.75 |
| 273 | 1481 | 0.81 | 0.88 | 1.48 |
| 253 | 1558 | 0.97 | 0.72 | 1.49 |

233 MHz

| T(K) | $\Delta v$ (Hz) | $T_1$ (s) | $T_2$ (ms) |
|---|---|---|---|
| 373 | 910 | 0.32 | 4.33 |
| 353 | 935 | 0.35 | 2.55 |
| 333 | 992 | 0.32 | 1.71 |
| 313 | 1072 | 0.33 | 0.91 |
| 298 | 1158 | 0.26 | 0.70 |
| 273 | 1281 | 0.32 | 0.51 |
| 253 | 1412 | 0.33 | 0.48 |
| 233 | 1595 | 0.26 | 0.38 |
| 213 | 1972 | 0.46 | 0.25 |
| 173 | 3937 | 1.44 | 0.38 |

Table 2. NMR spin-lattice relaxation time ($T_1$), spin-spin relaxation time ($T_2$), NMR linewidth ($\Delta v$) at 291 MHz and 233 MHz and the rotational frame of reference spin-lattice relaxation time $T_{1,\delta}$ at an effective frequency of 133 kHz for $Li_3YCl_6$.

291 MHz

| T(K) | $\Delta v$ (Hz) | $T_1$ (s) | $T_2$ (ms) | $T_{1,\delta}$ (ms), $f_{eff}$ = 133 kHz |
|---|---|---|---|---|
| 353 | 1681 | 0.043 | 0.29 | 0.41 |
| 343 | 1975 | 0.050 | 0.25 | 0.32 |
| 333 | 2502 | 0.061 | 0.22 | 0.25 |
| 323 | 2983 | 0.072 | 0.18 | 0.20 |
| 313 | 3577 | 0.086 | 0.15 | 0.17 |
| 303 | 4100 | 0.11 | 0.15 | 0.16 |
| 293 | 4722 | 0.12 | 0.12 | 0.17 |
| 273 | 5622 | 0.17 | 0.12 | 0.19 |
| 253 | 6357 | 0.23 | 0.11 | 0.29 |

233 MHz

| T(K) | $\Delta v$ (Hz) | $T_1$ (s) | $T_2$ (ms) |
|---|---|---|---|
| 403 | 409 | 0.015 | 0.65 |
| 373 | 637 | 0.019 | 0.30 |
| 353 | 1352 | 0.036 | |
| 333 | 1776 | 0.036 | |
| 298 | 3101 | 0.086 | |
| 278 | 3281 | 0.082 | |
| 273 | 3376 | 0.12 | |
| 253 | 3695 | 0.14 | |
| 233 | 4152 | 0.19 | |
| 213 | 5423 | 0.22 | |
| 193 | 6375 | 0.36 | |
| 173 | 7678 | 0.86 | |

Table 3. NMR spin-lattice relaxation time ($T_1$), spin-spin relaxation time ($T_2$), NMR linewidth ($\Delta v$) at 291 MHz and 233 MHz and the rotational frame of reference spin-lattice relaxation time $T_{1,\delta}$ at an effective frequency of 133 kHz for $Li_{1.48}Al_{0.48}Ge_{1.52}(PO_4)_3$.

291 MHz

| T(K) | $\Delta v$ (Hz) | $T_1$ (s) | $T_2$ (ms) | $T_{1,\delta}$ (ms), $f_{eff}$ = 133 kHz |
|---|---|---|---|---|
| 333 | 1665 | 0.27 | 4.4 | 8.50 |
| 323 | 1671 | 0.29 | 3.88 | 7.02 |
| 303 | 1782 | 0.43 | 1.77 | 4.11 |
| 293 | 1664 | 0.60 | 1.66 | 3.10 |
| 273 | 1701 | 1.00 | 0.90 | 3.02 |
| 253 | 1675 | 1.29 | 0.55 | 1.42 |

233 MHz

| T(K) | $\Delta v$ (Hz) | $T_1$ (s) | $T_2$ (ms) |
|---|---|---|---|
| 373 | 1126 | 0.22 | 7.59 |
| 333 | 1335 | 0.36 | 2.15 |
| 313 | 1210 | 0.43 | 2.08 |
| 298 | 1323 | 0.57 | 1.32 |
| 273 | 1409 | 0.79 | 1.09 |
| 233 | 1508 | 1.30 | 0.24 |
| 193 | 2014 | 1.44 | 0.21 |
| 173 | 2484 | 1.87 | 0.32 |

Table 4. NMR spin-lattice relaxation time ($T_1$), spin-spin relaxation time ($T_2$), NMR linewidth ($\Delta v$) at 291 MHz and 233 MHz and the rotational frame of reference spin-lattice relaxation time $T_{1,\delta}$ at an effective frequency of 133 kHz for $Li_6PS_5Cl$.

291 MHz

| T(K) | $\Delta v$ (Hz) | $T_1$ (s) | $T_2$ (ms) | $T_{1,\delta}$ (ms), $f_{eff}$ = 133 kHz |
|---|---|---|---|---|
| 293 | 651 | 0.42 | 10.38 | 11.15 |
| 303 | 660 | 0.35 | 14.72 | 16.69 |

233 MHz

| T(K) | $\Delta v$ (Hz) | $T_1$ (s) | $T_2$ (ms) |
|---|---|---|---|
| 403 | 589 | 0.39 | 45.17 |
| 383 | 581 | 0.29 | 41.29 |
| 363 | 582 | 0.24 | 37.46 |
| 343 | 588 | 0.22 | 28.79 |
| 323 | 602 | 0.23 | 23.13 |
| 313 | 600 | 0.24 | 19.71 |
| 303 | 601 | 0.27 | 15.17 |
| 298 | 604 | 0.28 | 12.66 |
| 293 | 585 | 0.30 | 9.97 |
| 288 | 612 | 0.36 | 6.99 |
| 273 | 670 | 0.46 | 3.66 |
| 253 | 727 | 0.62 | 1.33 |
| 233 | 751 | 1.01 | 0.64 |
| 213 | 814 | 1.18 | 0.38 |
| 193 | 1142 | 1.66 | 0.36 |
| 173 | 2131 | 2.45 | 0.37 |

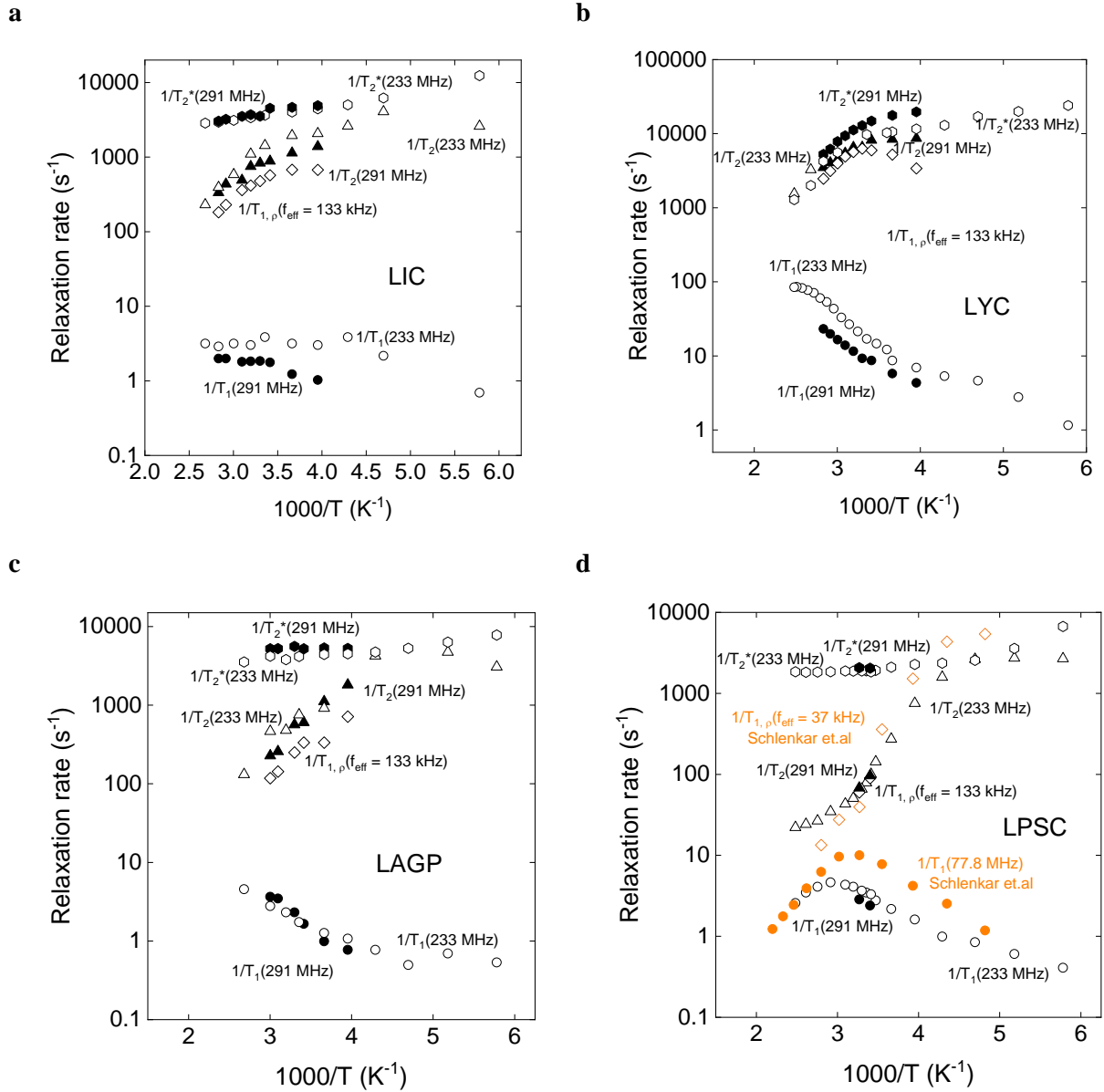

Figure 3. NMR relaxation rates for different values of 1000/T(K) for (a) LiInCl$_6$ (LIC), (b) LiYCl$_6$ (LYC), (c) Li$_{1.48}$Al$_{0.48}$Ge$_{1.52}$(PO$_4$)$_3$ (LAGP) and (d) Li$_6$PS$_5$Cl (LPSC). The data at 233 MHz are plotted as open symbols and at 291 MHz are plotted as filled symbols. The spin-lattice relaxation rate (1/$T_1$) are represented by circles, spin-spin relaxation rate (1/$T_2$) are represented by triangles and the rate of FID relaxation (1/$T_2$* = 2π(Δν)) are represented by hexagons. The spin-lattice relaxation rate at the effective rotational frame of reference frequency of 133 kHz is represented by open diamonds. In (d) the rates determined by Schlenker et.al[2] of spin-lattice relaxation rates at 77.8 MHz (orange filled circles) and at an effective field of 37 kHz (orange open diamonds) are included for comparison.

In the BPP model of dipolar interaction between two nuclei, the rate of spin-lattice relaxation is given by[2,6,9] $\frac{1}{T_1} = \frac{C\tau_c}{1+(\omega\tau_c)^{1+\beta}}$ where $C$ is a constant of dipolar interaction, $\tau_c$ is the correlation time of the dipolar interaction, and $\omega$ is the resonance frequency. The spin-spin relaxation time is given by[5]: $\frac{1}{T_2} = C\left[\frac{3}{2}\tau_c + \frac{5}{2}\frac{\tau_c}{1+\omega^2\tau_c^2} + \frac{\tau_c}{1+4\omega^2\tau_c^2}\right]$.

Therefore, according to BPP theory, the rate of spin-lattice relaxation reaches a maximum when the rate of atomic scale hopping, i.e., the inverse of the correlation time of Li-Li interaction, is equal to the resonance frequency $\omega$. In LPSC, we observe a spin-lattice relaxation around 343 K at 233 MHz, indicating, according to BPP theory, the rate of atomic scale hopping is comparable to $2\pi(233 \times 10^6)$/s at 343 K. In all other electrolytes, we do not observe a maximum in the $T_1$ relaxation in the temperature range of investigation at 233 MHz or 291 MHz.

The $T_{1,\delta}$ measurements at 133 kHz were carried out to understand the field dependence of $T_1$ relaxation, since the difference between 233 MHz and 291 MHz was not significant. However, in the range of temperatures available on the VNS 750 solid-state probe, $T_{1,\delta}$ is $T_2$-limited, and therefore, we cannot make conclusions on the field dependence on $T_1$ based on our rotational frame of reference $T_{1,\delta}$ measurements.

In LIC, LYC and LAGP the spin-spin relaxation times do not seem to have a field dependence, consistent with the BPP theory when $\frac{1}{\tau_c} \ll \omega$. For the limited data we have obtained for LPSC at 291 MHz at 293 K and 303 K, the $T_2$ values are similar to the values at 233 MHz for the same temperatures. The data obtained in LPSC at 233 MHz above 343 K shows the temperature dependence of $T_2$ decreases compared to the temperature dependence below 343 K. This is consistent with the BPP theory where there are additional terms contributing to the $T_2$ relaxation when the rate of atomic scale hopping is comparable to the resonance frequency of the experiment.

In all the battery electrolytes, the temperature dependence of $T_2$ is stronger than the temperature dependence of $T_1$ for most of the temperature range in the measurement. This would indicate in the BPP model a value of β: 0 < β < 1. In the BPP model, the expected field dependence of the spin-lattice relaxation rate when the relaxation rate is much smaller than the resonance frequency $\omega$ is $\frac{1}{T_1} \propto \frac{1}{\omega^{\beta+1}}$.

In LPSC, Schlenkar et.al[2] found β = 0.52. Our observations at 233 MHz and 291 MHz and comparison to the data by Schelnkar et.al at are consistent with the field dependence of ~ $\frac{1}{\omega^{1.52}}$ for $T_1$ relaxation. The consequent hopping rates are also consistent with the results from ionic conductivity measurements[2].

In LIC and LYC, the field dependence of the spin-lattice relaxation rate is even larger than $\frac{1}{\omega^2}$ while for LAGP there does not seem to be a field dependence.

Therefore, in the temperature and fields investigated, the BPP model for dipolar interaction between two nuclei consistently explains the NMR relaxation rates in LPSC and does not explain the NMR relaxations in LIC, LYC and LAGP.

**Conclusions**

Temperature dependence of NMR relaxation rates in $LiInCl_6$, $LiYCl_6$, $Li_{1.48}Al_{0.48}Ge_{1.52}(PO_4)_3$ and $Li_6PS_5Cl$ were studied from 173 K to 403 K at $^7Li$ resonance frequency of 233 MHz and from 253 K to 353 K at $^7Li$ resonance frequency of 291 MHz and an effective frequency of 133 KHz. The electrolytes were characterized by XRD, EIS and EDXRF to ensure expected crystallinity, ionic conductivity and stoichiometry. In the temperature range and frequencies studied, the relaxation rates in $Li_6PS_5Cl$ seem to be consistent with the Bloembergen-Pound-Purcell (BPP) theory for dipolar interaction of two Li nuclei while in $LiInCl_6$, $LiYCl_6$ and $Li_{1.48}Al_{0.48}Ge_{1.52}(PO_4)_3$ BPP theory is insufficient to explain the rates of NMR relaxation.

**Acknowledgements**

The authors thank Dr. Lingyang Zhu and Dr. Andre Sutrisno of the School of Chemical Science, University of Illinois at Urbana-Champaign for assistance with the NMR measurements. Part of the research was carried out at the Materials Research Laboratory at the University of Illinois. This research was partly supported by US Army CERL W9132T-19-2-0008 and Semiconductor Research Corporation (Task ID: 3044.001).